\documentstyle[12pt]{article}
\begin{document}
\def \be {\begin{equation}}
\def \ee {\end{equation}}

\begin{titlepage}

\title{Chaos in short-range spin glasses}
\author{F. Ritort$^{1),2)}$}

\maketitle

\begin{center}
{\small\baselineskip=14pt
1) Dipartimento di Fisica, Universit\`a di Roma
II, ''Tor Vergata'',\\Via della Ricerca Scientifica, Roma
00133, Italy}
\vskip 4pt
{\small\baselineskip=14pt
2) Departament de F\'{\i}sica Fonamental, Universitat de Barcelona,\\
Diagonal 647, 08028 Barcelona, Spain}

\end{center}

\vskip 10mm
\quad     Short title: Chaos in spin glasses
\vskip 10mm

\par
\quad PACS. 75.5 0 L-- Spin glasses.
\vskip .1in
\begin{abstract}
The nature of static chaos in spin glasses is studied. For the problem
of chaos with magnetic field, scaling relations in the case of the SK
model and short-range models are presented. Our results also suggest
that if there is de Almeida-Thouless line then it is similar to that
of mean-field theory for $d=4$ and close to the $h=0$ axis for $d=3$.
We estimate the lower critical dimension to be in the range $2.7-2.9$.
Numerical studies at $d=4$ show that there is chaos against
temperature changes and the correlation length diverges like $\xi\sim
(T-T^*)^{-1}$.
\end{abstract}
%\begin{flushright}
%{\bf Preprint ROM2F/93/ }
%\end{flushright}

\end{titlepage}

\baselineskip 6mm

\newpage

The nature of the spin-glass phase is poorly understood in short-range
spin glasses\cite{Bi86}. One very interesting topic is the static
chaos problem.  By this we mean how the free energy landscape is
modified when a small perturbation is applied to the system.

Most generally one is interested in the problem of chaos when a small
magnetic field is applied or when the temperature of the system is
slightly changed. The interest of this problem is twofold.  On the one
hand, it is important to discover what is the nature of the spin-glass
phase in short-range models. The study of chaos can give
interesting predictions regarding this question.  On the other
hand, it is relevant for the understanding of some recent cycling
temperature experiments in spin glasses \cite{Re87,Sa88}. Under the
hypothesis that in the dynamical experiment one is probing some kind
of equilibrium states \cite{Ha92} it is of the utmost importance to
understand the effect of changing the temperature on the free energy
landscape.

In this letter we shall present results on the problem of chaos in a
magnetic field and will see how scaling arguments may be used within
the spin-glass phase. From these arguments one can predict the
nature of the spin-glass phase in finite dimensions. Also in the
case when the temperature is changed we present some results which
show that in finite dimensions the system is more chaotic to
temperature changes than in mean-field theory.

The study of chaos with a magnetic field was adressed by I. Kondor in
case of mean-field theory \cite{Ko89}. Let us suppose two copies of
the same system (i.e. with the same realization of bonds): one at zero
magnetic field, the other one at finite magnetic field $h$ (in the
general case one could study different non-zero magnetic fields but
for simplicity we will focus on the particular case in which one of
the magnetic fields is zero.)

The hamiltonian can be written as:

 \be
H[\sigma\,,\tau]=-\sum_{i<j}\,J_{ij}\,\sigma_i\,\sigma_j
-\sum_{i<j}\,J_{ij}\,\tau_i\,\tau_j- h\sum_i\tau_i
\label{eq1}
\ee
One defines the order parameter
\be
q=\frac{1}{N}\,\sum_i\sigma_i\tau_i
\label{eq3}
\ee
and its corresponding correlation function
\be
C(i-j) =\overline{\langle\sigma_i\sigma_j\tau_i\tau_j\rangle}
\label{eq4}
\ee
where $\overline{(\cdot)}$ means average over disorder and
$\langle(\cdot)\rangle$ is the usual thermal average over the
Hamiltonian eq.(\ref{eq1}).

It has been shown by I. Kondor that this correlation function decays
to zero at large distances with a finite correlation length which
diverges for $h\to 0$. For a finite field $q=0$ is a stable solution
and the system is chaotic. In fact, the propagator $G(p)$ (i.e. the
Fourier transform of $C(i-j)$) can be exactly computed within the
Gaussian approximation.  Its singular part is given by \cite{Ko89}:
\be
G(p)=\int_{0}^{q_{max}}\,dq\,\int_{Q_{min}}^{Q_{max}}\,dQ
\frac{p^2+1+\lambda(q)\lambda(Q)}{(p^2+1-\lambda(q)\lambda(Q))^3}
\label{eq5}
\ee
with
\be
\lambda(q)=\beta(1-q_{max}+\int_{q}^{q_{max}}\,dq\,x(q))
\label{eq6}
\ee
where $\beta$ is the inverse of the temperature. The same expression
applies in the case of $\lambda(Q)$. Here $q(x)$ and $Q(x)$
are the order parameter functions for the spin glass at zero and $h$
field respectively.

It was found that the correlation length $\xi$ diverges like
$(1-\lambda(Q_{min}))^{-\frac{1}{2}}$. Close to $T_c=1$ we
have $Q_{min}\sim h^{\frac{2}{3}}$. This gives $\xi\sim
h^{-\frac{2}{3}}$. These are the results already obtained in reference
\cite{Ko89}.

Now we want to extract more information on the transition when $h\to 0$.
We can define a certain kind of non-linear susceptibility by:
\be
\chi_{nl}=\sum_i C(i)
\label{eq7}
\ee
It can be shown that it is also given by $\chi_{nl}=N\,\overline{\langle
q^2\rangle}$ with $q$ given in eq.(\ref{eq3}).  In terms of
eq.(\ref{eq5}) we have $\chi_{nl}=G(0)$. Using the known expressions
\cite{Pa80} for $q(x)$ and $Q(x)$ close to $T_c$ in eq.(\ref{eq5}) we
obtain a divergent expression for $G(0)$. Its most divergent part is
given by
\be
G(0)\sim
\int_{Q_{min}}^{Q_{max}}\,\frac{dQ}{(1-\lambda(Q))^{\frac{5}{2}}}
\label{eq8}
\ee
which gives $\chi_{nl}\sim\xi^4\sim h^{-\frac{8}{3}}$.  In the
Gaussian approximation this gives $G(x)\sim\frac{1}{x^{\mu}}$ for
$h\to 0$ with $\mu=d-4$. This means that correlation functions decay
more slowly in the spin-glass phase than they do at the critical point
($\mu=d-2$) and this is a consequence of the existence of a large
number of marginal states in the spin-glass phase.

In general we can introduce exponents $\gamma$ and $\nu$ in the
spin-glass phase such that $\chi_{nl}\sim h^{-\gamma}$ and $\xi\sim
h^{-\nu}$. Now we are interested in deriving some finite-size scaling
relations in order to establish (if it exists) the upper critical
dimension below which we expect scaling to be satisfied. Also scaling
relations can be very helpful to test predictions using numerical
simulations.

The derivation of scaling relations within the spin-glass phase
proceeds analogously as is done at the critical point. At the critical
point (below six dimensions) we can write $\chi_{nl} \sim
L^{2-\eta}\,f(N\,h^2\,q)$ because $h^2\,\sum_{a<b}\,Q_{ab}$ is the
singular part of the free energy per site
($Q_{ab}=\langle\sigma_a\sigma_b\rangle$ with $a,b$ replica indices).
At the critical point $(Q_{ab}=q)$ we introduce the exponent $\delta$
such that $q=h^{\frac{2}{\delta}}$. From these results one gets $\xi
\sim h^{-\frac{2(\delta +1)}{d\delta}}$ and the scaling relation
\be
\chi_{nl}=L^{2-\eta}\,f(L\,h^{\frac{2(\delta +1)}{d\delta}})
\label{eq10}
\ee
This gives the usual hyperscaling relations $\beta\,(\delta+1)=d\nu$ and
$\delta=\frac{d+2-\eta}{d-2+\eta}$. In mean-field theory \cite{Sk75} we
have $q\sim h$ ,i.e $\delta=2,\,\nu=\frac{1}{2}$. Hyperscaling relations
give $\eta=0$ and $d=6$ which is the upper critical dimension.

In the spin-glass phase, the derivation is slightly different because
$Q_{ab}\ne 0$. To obtain the singular part of the free energy one has to
substract from $h^2\,\sum_{a<b}\,Q_{ab}$ that part corresponding
to zero magnetic field.  In mean-field theory this is given by
$h^2(\int_{Q_{min}}^{Q_{max}}Q(x)dx\,-\,\int_{0}^{q_{max}}q(x)dx)$ which
is proportional to $h^2\,Q_{min}\,x_{min}$ with $x_{min}$ equal to the
first breakpoint of the function $Q(x)$. Because $Q_{min}\sim
x_{min}\sim h^{\frac{2}{3}}$ we obtain
\be
\chi_{nl}=\,N^{\frac{4}{5}}\,f(N\,h^{\frac{10}{3}})
\label{eq11}
\ee
If an upper critical dimension exists then it should be $d_u=5$
because for that dimension $\xi\sim\,h^{-\frac{2}{3}}$.

We have performed Monte Carlo numerical simulations of the SK model in
order to test the prediction eq.(\ref{eq11}). The results for
different magnetic fields ranging from $0.2$ up to $1.0$ at $T=0.6$
are shown in figure 1 for several sizes (up to $N=2016$). There is
agreement with the prediction eq.(\ref{eq11}) and $\chi_{nl}$ versus
the field $h$ does not change its behaviour when crossing the de
Almeida-Thouless line ($h_{AT}\sim 0.45$ for $T=0.6$).

Now we present the results of our simulations and our predictions for
short-range models. We expect that scaling is satisfied in the
spin-glass phase below five dimensions. In general, one has:
\be
\chi_{nl}=L^{\lambda}\,f(L^d\,h^2\,Q_{min}\,x_{min})
\label{eq12}
\ee

The critical point is a particular case of eq.(\ref{eq12}). Since
there is not replica symmetry breaking one has $x_{min}\sim 1$ and
$Q_{min}=q\sim h^{\frac{2}{\delta}}$. Putting $\lambda=2-\eta$ one
recovers eq.(\ref{eq10}).  To test this expression at the critical
point we have performed Monte Carlo numerical simulations of the $4d$
$\pm J$ Ising spin glass (with periodic boundary conditions) which is
known to have a transition at $T=2.06\pm0.02$ \cite{Ba93,Si86}. In
this case the known values $\eta\simeq -0.25,\,\delta\simeq 3.6$ fit
the data reasonably well but even though the scaling is very sensitive
to the precise value of the critical temperature. This shows that, at
the critical point, the critical behavior of eq.(\ref{eq4}) is the
same as that of the correlation function of two identical copies of
the system at the same field $h$.

Now we present our results at $T=1.5$ in the spin-glass phase for the
$4d$ Ising spin glass. One has to be sure that samples are well
thermalized. To this end we have performed a simmulated annealing
algorithm which reaches equilibrium in a reasonable time. Our results
in figure 2 fit well a scaling law $\chi_{nl}\sim
L^{3.25}\,f(h\,L^{1.45})$.  This gives $\lambda\simeq 3.25$,
$\nu\simeq 0.69$ and $\gamma=\lambda\nu\simeq 2.24$. We should draw
attention to the fact the value found for $\lambda$ is close to
that found in mean-field theory eq.(\ref{eq11}) putting $N=L^4$.

We have also studied the $3d$ $\pm J$ Ising spin glass which has a
transition close to $T=1.2$ \cite{Bh88,Og85}. Simulations in the
critical temperature give exponents in eq.(\ref{eq10}) in agreement
with those already known. We have also performed simulations for small
sizes at $T=0.8$ in the spin-glass phase. Our results are compatible
with $\lambda\sim 2.4$, $\nu\sim 0.7$ and $\gamma=\lambda\nu\simeq
1.9$

{}From our results at $d=3,4$ it seems that $\lambda=\frac{4d}{5}$ and
$\nu=\frac{2}{3}$ are a good approximation to the exponents at least for
$d\leq 5$.

Now we can adress the question of the existence of a phase transition
line in finite magnetic field in the $4d$ Ising spin glass (the so
called AT line \cite{At78}). From a theoretical point of view the
problem remains unsolved \cite{Br80}. Recent numerical studies suggest
that this line really exists \cite{Gr92,Ba93,Ri93}. If this is the
case and there is also mean-field behaviour in the spin-glass phase at
zero magnetic field \cite{Pa93,Ci93} then it is natural to suppose
that (as happens in mean-field theory) $Q_{min}\sim x_{min}$.  This is
a very plausible hypothesis which agrees with the fact that $q(x)\sim
x$ for small $x$, or that $P(0)=(\frac{dx}{dq})_{q=0}$ has a finite
value \cite{Re90}.  If $Q_{min}\sim h^{\frac{2}{\delta}}$ ($\delta=3$
in the mean-field case) we obtain, from eq.(\ref{eq12}) the result
$\xi\sim h^{-\frac{2(2+\delta)}{d\delta}}$. For $d=4$ this gives
$\delta\simeq 5.27$ and $Q_{min}\sim h^{0.38}$. In three dimensions
$\delta$ is uncertain but is a very large value (of order
$20$).

We expect the transition line in magnetic field to occur when
$Q_{min}\sim Q_{max}$, i.e. $h^{\frac{2}{\delta}}\sim \tau^{\beta}$ or
$h\sim\tau^{\frac{\beta\delta}{2}}$ with $\tau=T_c-T$.  Because
$\beta\sim 0.6,0.5$ in $d=4\cite{Pa93},3\cite{Og85}$ respectively this
gives $h\sim\tau^{1.58}$ in four dimensions and $h\sim\tau^r$ with $r$
of order five in three dimensions. This means that in three dimensions
the AT line is very close to the $h=0$ axis and very difficult to see
numerically, at least not very far from $T_c$ \cite{Ca90,Ka92}.

{}From the scaling relation eq.(\ref{eq12}) we can also estimate the
lower critical dimension. If we define the exponent $\theta$ as that
exponent for which $(\xi^d h^2 q)^{\frac{1}{2}}\sim \xi^{\theta}$ where
$q$ is finite then this gives the thermal exponent introduced in droplet
models \cite{Br87,Ko88,Fi88}. The exponent $\theta$ vanishes for $d=d_l$
where $d_l$ is the lower critical dimension. We have obtained
$\theta=1,0.55,0.05$ in $d=5,4,3$ respectively. Extrapolating to
$\theta=0$ we estimate $d_l=2.7-2.9$ which is in agreement with
perturbative calculations in the spin-glass phase \cite{Do92} but higher
than the value reported in \cite{Ni92}.

We have also investigated the problem of chaos against temperature
changes. The outline of the ideas follow that presented above in the
case of a magnetic field. Now one couples two copies of the system at
different temperatures. In mean-field theory the problem has not yet
been fully solved and chaos could be marginal
\cite{Ko89,Fr93}. In finite dimensions a interesting behaviour is
expected in low dimensions \cite{Ni92}. Our numerical results for the SK
model indicate that, if there is chaos, it is very small (details will
be presented elsewhere). We have performed simulations in the $4d$
Ising spin glass.Figure 4 shows the non-linear susceptibility defined in
eq.(\ref{eq7}) using eq.(\ref{eq3}) which is the overlap obtained by
coupling two identical copies of the system at different temperatures.
Our results are consistent with a correlation length which diverges like
$(T-T^*)^{-1}$ where $T^*$ is the reference temperature of one of the
two copies. This results are in agreement with perturbative calculations
in the range of dimensionalities $6<d<8$ \cite{Ko93}.

Summarizing, in the case of chaos with a magnetic field we find that
there is scaling behaviour in the spin-glass phase in mean-field theory,
the main result being eq.(\ref{eq8}). Short-range systems also satisfy
scaling relations from which we can extract the exponents associated to
the correlation length. We derive that if there is an AT line then in
$d=4$ it is similar to that of mean-field theory and in $d=3$ it is very
close to the $h=0$ axis and more difficult to see using numerical
simulations.  The lower critical dimension is also predicted to be in
the range $d_l\sim 2.7-2.9$. We also reported some results of chaos in
temperature which show that short-range models are more chaotic than the
SK model and the correlation length diverges like $(T-T^*)^{-1}$.

\paragraph{Acknowledgements}

I gartefully acknowledge very stimulating conversations with G.
Parisi, A. J. Bray, M. A. Moore and S. Franz. This work has been
supported by the European Community (Contract B/SC1*/915198).

\vfill\eject

\vfill\eject
\begin{center}
FIGURE CAPTION
\end{center}

\vskip1truecm

Fig. 1 \quad Chaos with magnetic field in the SK model at $T=0.6$. Field
values range from $h=0.2$ up to $h=1.0$ for the smaller sizes and up to
$h=0.4$ for the largest ones.The number of samples range from $200$ for
$N=32$ down to $25$ for $N=2016$.
\vskip1truecm
Fig. 2 \quad Chaos with magnetic field in the $4d$ Ising spin glass at
$T=1.5$. Magnetic field values range from $h=0.1$ up to $h=1.$. The
number of samples is approximately $100$ for all lattice sizes.
\vskip1truecm
Fig. 3 \quad Chaos with temperature changes in the $4d$ Ising spin
glass. The reference temperature of one copy is $T^*=1.5$. Temperature
values of the other copy range from $1.6$ up to $2.2$. The number of
samples is approximately $100$ for all lattice sizes.

\vfill\eject

\end{document}